\documentstyle[preprint,aps]{revtex}

\def\q{\quad}
\def\bD{\bar{D}}
\def\dalpha{{\dot{\alpha}}}
\def\btheta{\bar{\theta}}
\def\sg{\sigma}
\def\lrw{\longleftrightarrow}
\newcommand{\bam}{\left( \begin{array}}
\newcommand{\eam}{\end{array} \right)}
\newcommand{\bamq}[4]{\left( \begin{array}{cccc}{#1}&{#2}&{#3}&{#4}\\}

\begin{document}
\draft
\preprint{Lecce University}
\title{
Octonions
and  Super Lie Algebra}
\author{
Khaled Abdel-Khalek\footnote{Work supported by an
ICSC--World Laboratory scholarship.\\
e-mail : khaled@le.infn.it}
}
\address{ 
Dipartimento di Fisica - Universit\`a di Lecce\\
- Lecce, 73100, Italy -}

\date{Jan. 1997}
\maketitle
\begin{abstract}
We  discuss how to represent the non-associative
octonionic structure in terms of the associative matrix algebra
using the left and right octonionic operators.
As an example
we construct explicitly some Lie and Super Lie algebra.
Then we discuss the notion of octonionic Grassmann  
numbers and explain its possible application 
for giving a superspace formulation of 
the minimal supersymmetric Yang-Mills  models
.
\end{abstract}

\widetext
\vspace{2cm}

Usually we define an almost complex manifold as a real
manifold equipped with a complex structure ${\cal I}$ such
that ${\cal I}^2 = -1$ which may  be 
a matrix like  \[ \left( \begin{array}{cc}
0&-1\\ 1&0 \end{array}\right) \] and the same holds equally well
for a quaternionic manifold but we would have
${\cal I, J, K}$ respecting an $su(2)$ algebra. Generalizing
this notion to octonions we meet a puzzle, how to represent
the non-associative structure of octonions. Can it be only
in terms of some defined operators or can we find an easy
way to do it with matrices? the answer is indeed yes, we can again do
it with matrices but we need a trick. 

We  use the  
symbols $e_i$ to denote the imaginary octonionic units
where $i,j,k = 1 .. 7$ and $e_i. e_j = -\delta_{ij} + \epsilon_{i j k} e_k$
or $[ e_i , e_j ] =
\epsilon_{ijk} e_k$ such that 
$\epsilon_{i j k}$ equals 1 for one of the following 
seven combinations \{(123),(145),(176),(246),(257),(347),(365)\}. 

We know that, from a topological point of view,
any ${\cal R}^{8}$ is a trivial octonionic manifold.
So we can represent an octonion as  8 dimensions column
matrix but as octonions are non-commutative and non-associative
different action from right or left and taking into account
their peculiar non-associativity property may give rise
to 106 left/right operators which may be constructed
completely from the following 14 operators
$\{ E_1 , ... , E_7 , 1|E_1 , ... , 1|E_7 \}$
\cite{my1,schaf}, we mean by $e_i$ an octonionic number whereas
 $E_i$ are
their corresponding  
matrices and $1|E_i$
represent action from right,
i.e 
they are the corresponding matrix form of $1|e_i$ 
given by\footnote{We  use the elegant 
notations of \cite{rotl,rot1}.}
\begin{eqnarray}
1|e_i~~g = g~~e_i \quad \quad \quad g \in {\cal O}.
\end{eqnarray}
We have given a matrix as well as a tensorial representation
of these fundamental 14 operators in a separate appendix.

One can check explicitly that
any of these matrices square to -1
but they don't obey the octononic multiplication
table. 
\begin{eqnarray}
( E_{i} )^2 &=&  - \openone ,\label{w1}\\
(~1|E_{i} )^2 &=& - \openone ,\label{w2}\\
E_i ~~1|E_i &=& 1|E_i ~~E_i ,\\
\{E_i , E_j \} &=& -2 \delta_{ij} \openone ,\label{d1}\\
\{ 1|E_i , 1|E_j \} &=& -2 \delta_{ij} \openone ,\label{d2}\\
~[ E_i , E_j ] &=&  2 \epsilon_{ijk} E_k - 2 [ E_i, 1|E_j ] ,\label{si1}\\
~[ 1|E_i , 1|E_j ] &=& 2 \epsilon_{jik}~1|E_k - 2 [ E_i, 1|E_j ] .\label{si2} 
\end{eqnarray}
Moreover, any of the left or right set alone
doesn't close an algebra but only 
when we allow their mixing, we can get something useful.
To this moment we should recall how an 
octonionic structure is usually constructed. 
In any work about octonions, one extracts 
the octonionic structure by  geometric meaning from
spaces with $SO(8)$ or $SO(7)$ holonomy.
The argument is simple any unit octonion is
isomorphic to the Reimannian $S_7 \sim SO(8)/SO(7)$ or to any of its
homeomorphic squashed versions $S_7^{\prime} \sim
SO(7)/G_2$ or $S_7^{\prime \prime} \sim SU(4)/SU(3)$ and lastly
$S_7^{\prime \prime \prime} \sim Sp(2)/Sp(1)$ so having
a manifold with holonomy group $SO(8), SO(7), SU(4)\sim SO(6)
$
, or even $G_2$, the octonionic automorphism group, 
one can in principle extract the octonionic structure.
But any of these groups ($SO(8), SO(7), SO(6), SO(5), G_2$) 
admits an explicit construction using the left and right 
$E_i, 1|E_i$ \cite{choq}. So they represent the ``associativizing''
form of the non-associative octonionic imaginary units $e_i, 1|e_i$.
Simply 
\begin{eqnarray}
\{\frac{1}{2} E_i , \frac{1}{4} [ E_i , E_j ] \}
\quad i,j = 1 .. 7 \quad \mbox{close  Spin(8)} \label{a1}\\
\{ \frac{1}{4} [ E_i , E_j ] \}
\quad i,j = 1 .. 7 \quad \mbox{close  Spin(7)} \\
\{ \frac{1}{4} [ E_i , E_j ] \}
\quad i,j = 1 .. 6 \quad \mbox{close  Spin(6)} \\
\{ \frac{1}{4} [ E_i , E_j ] \}
\quad i,j = 1 .. 5 \quad \mbox{close  Spin(5)} \\ 
\{ \frac{1}{4} [ E_i , E_j ] \}
\quad i,j = 1 .. 4 \quad \mbox{close  Spin(4)} \\ 
\{ \frac{1}{4} [ E_i , E_j ] \}
\quad i,j = 1 .. 3 \quad \mbox{close  Spin(3)} \label{a2} 
\end{eqnarray}
and the same construction
can be done using the $\{1|E_i\}$ set. 
For further study of $G_2$ look at \cite{gg} 
. Actually the logic behind this construction is very
easy, upon the use of (\ref{w1}, \ref{w2},
\ref{d1}, \ref{d2})
, it is easy to see  that the two sets $\{E_1 , ... , E_7 \}$
and $\{1|E_1 , ... , 1|E_7\}$ generate Clifford Algebra
Cliff(0,7) then all the above given construction follows
except for $Spin(8)$ which follows from $SO(8) \sim
S_7 \times SO(7)$.

In summary, our philosophy is : these matrices can be used to 
investigate/detect
octonions easily using matrices. The non-associativity will be
represented by the non closure of the algebra.
Our 
left/right 14 operators, $\{E_1 , ... , E_7 , 1|E_1 , ... ,
 1|E_7 \}$  satisfy the Jacobi identity but they
don't close an algebra. To close an algebra
we should allow their mixing as it is clear from 
(\ref{a1})--(\ref{a2}) whereas octonions close an algebra
but they don't satisfy the Jaccobi identity.

One may even go further and check if these matrices admit 
a  Super Lie algebra (SLA).
We know that any Super Lie Algebra  is defined as a vector space which is 
the union of an even (bosonic) part B and another
odd (fermionic) part F such that
\begin{eqnarray}
~\{ F, F \} \in B , \\
~[ B, B ] \in B, \\
~[ F, B ] \in F ,
\end{eqnarray}
and we have the following four Super Jacobi Identities (SJI)
\begin{eqnarray}
& &~[~a,~[~b,c~]~]~ 
+ ~[~b,~[~c,a~]~]~ + ~[~c,~[~a,b~]~]~ = 0 , \quad \quad
a,b,c\in B , \label{s1}\\
& & ~[~a,~[~b,c~]~]~ 
+ ~[~b,~[~c,a~]~]~ + ~[~c,~[~a,b~]~]~ = 0 , \quad \quad
a,b\in B \ and\ c\in F , \label{prob}\\
& & ~[~a,~\{~b,c~\}~]~ 
+ ~\{~b,~[~c,a~]~\}~ - ~\{~c,~[~a,b~]~\}~ = 0 , \quad \quad
a\in B\ and\ b,c \in F , \\
& & ~[~a,~\{~b,c~\}~]~ 
+ ~[~b,~\{~c,a~\}~]~ + ~[~c,~\{~a,b~\}~]~ = 0 , \quad \quad
a,b,c\in F \label{s2},
\end{eqnarray}

Amazingly enough three of these SJI are  satisfied by
the octonionic elements $e_i$ under the following
decomposition
\begin{eqnarray}
  B=\{
e_1,e_2,e_3\} \quad and \quad 
F=\{e_0,e_4,e_5,e_6,e_7\} ,
\end{eqnarray}
problems arise because
of (\ref{prob}) which is the true reflection of the 
non-associativity, but (\ref{s1})--(\ref{s2}) are 
satisfied by the matrices $E_i , 1|E_i$ for
\begin{eqnarray}
  B_1 = \{
E_1,E_2,E_3\} \quad and \quad 
F_1 = \{\openone,E_4,E_5,E_6,E_7\} ,
\end{eqnarray}
 and the same for the right combination
\begin{eqnarray}
  B_2 = \{
1|E_1,1|E_2,1|E_3\} \quad and \quad 
F_2 = \{\openone,1|E_4,1|E_5,1|E_6,1|E_7\}.
 \end{eqnarray}
Using (\ref{d1}) and (\ref{d2}), one can check easily that
$F_1$ or $F_2$ close a  fermionic algebra, But
our bosonic algebra should be modified to
\begin{eqnarray}
B^{\prime}_1 = \{  [ E_1,E_2 ], [ E_1,E_3 ], [ E_2, E_3 ] \} ,\\
B^{\prime}_2 =  \{  [ 1|E_1, 1|E_2 ], 
[ 1|E_1, 1|E_3 ], [ 1|E_2, 1|E_3 ] \} ,
\end{eqnarray} 
using (\ref{si1}) and (\ref{si2}), it is obvious that
$B^{\prime}_1$ and $B^{\prime}_2$ close properly
under the commutation relation and generate an $so(3)\sim su(2)$ algebra.
It remains
to check, for example,
$[F_1,B^{\prime}_1] \in F$, to appreciate how difficult to work
analytically, let's try to use 
(\ref{si1}) and (\ref{si2}), for (we will work with
the $\{ E_i \}$ set but everything holds well for
the $\{ 1|E_i \}$ )
\begin{eqnarray}
[E_4 , [E_1,E_2] ] &=&
E_4 ( 2 E_3 - 2 [E_1 , (1|E_2) ] ) 
- ( 2 E_3 - 2 [E_1 , (1|E_2) ] ) E_4 
\nonumber \\
&=& 2 [ E_4, E_3 ] - 2 [ E_4 , [ E_1, (1|E_2) ] ] 
\nonumber \\ &=& 4 E_7 - 4 [ E_4, (1|E_3) ] - 2 [ E_4 , [ E_1, (1|E_2) ] ] 
,
\end{eqnarray}
we should prove that the last equation belongs to $F$,
one may even try to work with the tensorial notation
given in the appendix and invoke some octonionic identities
to find the answer. But, we have an easy way,
simply, we used the matrix representation given in the appendix
and find \footnote{One can use any computer system and
after entering the matrices given in the appendix, all what he
has to do is to define the commutator
and then he can check the
next equation.}
\begin{eqnarray}
[F_1, B^{\prime}_1 ] = [ F_2, B^{\prime}_2 ]
 = 0. \label{a28}
\end{eqnarray}
Now using  (\ref{w1}--\ref{d2}) and (\ref{a28}) 
the four SJI follow directly. So 
$\{B^{\prime}_1,F_1\}$ or $\{B^{\prime}_2,F_2\}$ 
is an SLA composed of a bosonic $su(2)$ and a 
four dimensional fermionic part. Even, we can generalize the
previous construction to
include
all the bosonic algebra $so(3)\sim su(2)$ to 
$so(6) \sim su(4)$
\begin{eqnarray}
B_6 &=& 
\{ \frac{1}{4} [ E_i , E_j ] \}
\quad i,j = 1 .. 6 \quad \mbox{close  Spin(6)} \\
F_6 &=& \{ \openone,E_7 \}.
\end{eqnarray}

Whatever we believe that the correct niche, for octonions, is local
supersymmetry, it seems to be a good idea to try, first,
something easier like global higher dimensional supersymmetry.
But we should first solve the following puzzle.
In trying to write supersymmetry over quaternions or octonions,
one faces the following difficulty. What is the correct definition
of a quaternionic or an octonionic Grassmann variables?
Usually, a Grassmann 
Algebra (GA) is defined by the following relation :
$~\{\theta_i,\theta_j\}~=~0$.
We want to show that this relation holds for quaternionic
or octonionic Grassmann variables without any 
modification. Actually, this should
 be anticipated
from the start as Grassmann variables are nothing but fermions.
Quaternionizing or octonionizing fermions is nothing but 
writing a quaternionic or  an octonionic representation of the 
corresponding Clifford algebra with 
a reduction of the number of components
of the spinor.

Grassmann numbers are defined as the set of anticommuting numbers
$\{\theta_1, 
\theta_2, ... ,\theta_n\}$ such that
$\forall i,j~=~1 ... n$
\begin{eqnarray}
\{ \theta_i , \theta_j \} = 0, \\
\theta_i^2 = 0. \end{eqnarray}
Whereas Cliff(p,q) 
is defined as the set of
$\{\gamma_1, ... , \gamma_p, \gamma_{p+1} , ... , \gamma_{p+q}\}$  
satisfying the following anticommutation relations
$(\forall n,m~=~1 ... p+q)$ and $\eta_{nm} \equiv 
(\overbrace{1,...,1}^{p}\overbrace{-1,...,-1}^{q})$
\begin{eqnarray} 
\{ \gamma_n, \gamma_m\} = 2 \eta_{nm}, 
\end{eqnarray}
For simplicity, consider $p=q$,
because of the signature, we will have
\begin{eqnarray}
\gamma_1^2 = \ldots =  \gamma_p^2 = 1\ \mbox{and}\ \ 
\gamma_{p+1}^2 = \ldots = \gamma_{2p} = -1.
\end{eqnarray}
then by coupling two elements of different signature, we have
\begin{eqnarray} 
(\gamma_1 &+& \gamma_{p+1})^2 = 0 , \nonumber\\
&\vdots& \nonumber \\
(\gamma_p &+& \gamma_{2p})^2 = 0 , \nonumber \\
\end{eqnarray}
i.e p Grassmann variables.
It is evident that we can not construct more than p-Grassmann variables.
To see this explicitly, consider p=2, we have
\begin{eqnarray}
\gamma_0, \gamma_1, i\gamma_2, \gamma_3 \label{aaa}
\end{eqnarray} 
where $(\gamma_i)$ are the standard Cliff(2,2) 
for example in the Dirac representation 
\[ \gamma_0^2 = (i\gamma_2)^2 = 1\ \ \mbox{whereas}\ \  
\gamma_1^2 = \gamma_3^2 = -1.
 \]
Then our Grassmann variables are nothing but
\[ \theta_1 = \gamma_0 + \gamma_1 \ \ \mbox{and}\ \
 \theta_2 = 
i\gamma_2 + \gamma_3.
\]
If one tries to introduce a third Grassmann variables as
\[ \theta_3 = \gamma_0 + \gamma_3 \]
we have the following situation
\[ \{ \theta_1, \theta_2 \} = 0, 
\{\theta_2, \theta_3 \} \neq 0 \ \mbox{and}\
\{\theta_1, \theta_3 \} \neq 0. \]
And generally for any 2p-dimensional Cliff.
of signature (p,p), one can construct p-Grassmann variables. 
It seems really that Clifford algebra is too fundamental.
It would be fantastic if the above construction
can be extended somehow to give the exact number of Grassmann variables
needed for the construction of supersymmetric theories.

Actually, the relation between supersymmetry
and ring division algebra is very clear in the construction of 
the minimal supersymmetric Yang-Mills theory 
\cite{evan,evan1,evan2,berk1,berk2}.
For example representing the D=6, 10 Lorentz group
as $sl(2, {\cal Q})$ and $sl(2, {\cal O})$ respectively
\cite{kugo},
then they admit the natural D=4, N=1 superspace 
construction as a solution.
A more general solution my be generated because  the 
Taylor expansion (used for example to find
the solution of the chiral field constraint) 
is not well defined at the level of 
a quaternionc or an octonionic formulation 
. In principle left and right as well as  their
mixing is allowed, explicitly (consult \cite{bag} for
notations) 
\begin{equation}
\bD_\dalpha \Phi \; = \; 0\;.
\label{dphi}
\end{equation}
This defines the chiral superfield, $\Phi$.  
We can solve the constraint (\ref{dphi}) by writing $\Phi$ as
a function of $y$ and $\theta$, where
\begin{equation}
y^m \; = \; x^m \; + \; i \theta \sigma^m \btheta\;.
\end{equation}
Since $\bD \theta = \bD y= 0$, the field $\Phi(y,\theta)$
automatically satisfies the constraint (\ref{dphi}).

To find the component fields, we expand $\Phi(y,\theta)$ in
terms of $\theta$,
\begin{eqnarray}
\Phi(y, \theta) & = & A(y) \; + \; \sqrt{2} \, \theta \chi(y) \;
+ \; \theta\theta \, F(y) \nonumber \\[2mm]
& = & A(x) \; + \; i \theta \sigma^m \btheta \, \partial_m A(x) \; 
+ \;{1\over4}\, \theta \theta \btheta \btheta \Box A(x) \nonumber \\
&& + \; \sqrt{2} \, \theta \chi(x) \; - \; {i\over\sqrt 2}\, \theta\theta \,
\partial_m \chi(x) \sigma^m \btheta \; + \; \theta \theta \, F(x)\;.
\label{phi expansion}
\end{eqnarray}
The problem is simply the following :  quaternionic or
 octonionic A(x)
doesn't commute any more with the quaternionic or octonionic
$\sigma^m$ i.e 
\begin{eqnarray}
i \theta \sigma^m \btheta \, \partial_m A(x)   
\neq  
i \partial^m A(x)   \theta \sigma^m \btheta \, , 
\end{eqnarray}
also, it is better to use the momentum operator $P^m$
instead of $i \partial^m$ 
, this is a technical problem related to
the quantization process of a quaternionic or octonionic fields, look in
\cite{my2} for more details \footnote{ Simply, adopting a complex scalar product
, the quantization process is the same. Adler \cite{Adl} goes 
further and proposes
that a complex scalar quantum mechanics has the same Hilbert space as
the standard quantum theory. From our point of view, this argument is not
clear since, even after the use of complex scalar, the theory still 
carries a quaternionic or an octonionic structure.}.
We think that this problem may be related
to the construction of the off-shell formulation
of the ten dimensional super Yang-Mills.

We think that supergravitational (gauged supersymmetric) theories
can be a good candidate for an octonionic gauge theory since
they are  torsionfull
version of the general relativity with
specific conditions imposed on the torsion tensor  
by the action of the Bianchi identities and it is also well known
that any  octonionic manifold
is a full torsion space. When we compactify
the simple N=1 D=10 super Yang-Mills 
to 4 dimensional, we get an N=4 SU(4) super Yang-Mills
; where the $SU(4)\sim SO(6)$ represents the
remnant of the higher dimensional Lorentz group.
By the same token, When we compactify on $S_7$ 
the D=11 N=1 supergravity
to 4 dimensions, we should look to the resultant theory, namely
the N=8 D=4 supergravity, as a full gauged $S_7$ theory.
To prove such conjectures, we should construct explicitly
the octonionic version of the D=4 N=8 or D=11 N=1
supergravity.

Lastly, octonion is a consistent wonderful part of mathematics and
finding their correct physical application, from our
point of view, is highly needed rather than just being  a 
challenge or a conjecture. In \cite{ced}, it has
been proposed to use the fact that octonions are 
``almost Lie algebra'' i.e locally, they close
a Lie algebra with the structure constant being a function 
of the coordinate. Unfortunately, one can not have a topological 
support of this notion, by applying different Hopf fibrations
\begin{eqnarray}
S_7 \longrightarrow S_4 \times S_3
\longrightarrow S_4 \times S_2 \times S_1 ,
\end{eqnarray}
also, this localization may miss some important
global features. As 
\begin{eqnarray}
\pi_7(S_7) \neq \pi_7(S_4 \times S_2 \times S_1).
\end{eqnarray}
Until finding
the correct way, one may try every possible physical/mathematical
formulation keeping in mind that our job as physicists
is to try our best to describe nature not to choose it.

\vspace{2cm}
 
I would like to acknowledge 
 P. Rotelli as well as the physics department
at Lecce university for their kind hospitality.
Also, I am grateful to Prof. A.~Zichichi 
and the ICSC--World Laboratory for financial
support. Last but not least, I would like to thank G. Thompson
for teaching me many beautiful topological notions through
my mathematical instantonic thesis.

\newpage
\appendix
\section*{}

We introduce the following notation:

\begin{eqnarray}
\{~a, \; b, \; c, \; d~\}_{(1)} ~ &\equiv& ~\bamq{a}{0}{0}{0} 0 & b & 0 & 0\\
0 & 0 & c & 0\\ 0 & 0 & 0 & d \eam    \q , \\ 
\{~a, \; b, \; c, \; d~\}_{(2)} ~ &\equiv& ~\bamq{0}{a}{0}{0} b & 0 & 0 & 0\\
0 & 0 & 0 & c\\ 0 & 0 & d & 0 \eam \q ,\\
\{~a, \; b, \; c, \; d~\}_{(3)} ~ &\equiv& ~\bamq{0}{0}{a}{0} 0 & 0 & 0 & b\\
c & 0 & 0 & 0\\ 0 & d & 0 & 0 \eam    \q , \\
\{~a, \; b, \; c, \; d~\}_{(4)} ~ &\equiv& ~\bamq{0}{0}{0}{a} 0 & 0 & b & 0\\
0 & c & 0 & 0\\ d & 0 & 0 & 0 \eam \q ,
\end{eqnarray}
where $a, \; b, \; c, \; d$ and $0$ represent $2\times 2$ real matrices.

In the following  $\sg_{1}$, $\sg_{2}$, $\sg_{3}$ represent the 
standard Pauli matrices.

\begin{eqnarray}
\begin{array}{ccccccc}
e_{1}  &\lrw&\{-i\sg_{2}, -i\sg_{2}, -i\sg_{2},  
i\sg_{2} ~\}_{(1)} \q &,&
1\mid e_{1} &\lrw&\{-i\sg_{2},  i\sg_{2}, i\sg_{2},  
-i\sg_{2} ~\}_{(1)}\q , \\
e_{2}  &\lrw&\{ -\sg_{3}, \sg_{3}, -1, 1 ~\}_{(2)}\q &,&
1\mid e_{2}&\lrw&\{ -1, 1, 1,  
-1 ~\}_{(2)}\q , \\
e_{3} &\lrw&\{ -\sg_{1}, \sg_{1}, -i\sg_{2},  
-i\sg_{2} ~\}_{(2)}\q &,&
1\mid e_{3} &\lrw&\{ -i\sg_{2}, -i\sg_{2}, i\sg_{2},  
i\sg_{2} ~\}_{(2)}\q , \\
e_{4} &\lrw&\{ -\sg_{3}, 1, \sg_{3}, -1 ~\}_{(3)}\q &,&
1\mid e_{4} &\lrw&\{ -1, -1, 1,  
1 ~\}_{(3)}\q , \\
e_{5} &\lrw&\{ -\sg_{1}, i\sg_{2}, \sg_{1},  
i\sg_{2} ~\}_{(3)}\q &,&
1\mid e_{5} &\lrw&\{ -i\sg_{2}, -i\sg_{2}, 
-i\sg_{2},  
-i\sg_{2} ~\}_{(3)}\q , \\
e_{6}&\lrw&\{ -1, -\sg_{3}, \sg_{3}, 1 ~\}_{(4)}\q &,&
1\mid e_{6} &\lrw&\{ -\sg_{3}, \sg_{3}, -\sg_{3},  
\sg_{3} ~\}_{(4)}\q , \\
e_{7} &\lrw&\{ -i\sg_{2}, -\sg_{1}, \sg_{1},  
-i\sg_{2} ~\}_{(4)}\q &,&
1\mid e_{7}&\lrw&\{ -\sg_{1}, \sg_{1}, -\sg_{1},  
\sg_{1} ~\}_{(4)}\q .
\end{array}
\end{eqnarray}

Following \cite{ok}
, It is easy to realize that our matrices, in 
tensorial notation,
are anti-hermitian
\begin{eqnarray}
<l|E_i|k> = - <k|E_i|l> ,
\end{eqnarray}
moreover,
\begin{eqnarray}
<0|E_i|k> = - <k|E_i|0> = - \delta_{ik},
\end{eqnarray}
and finally
\begin{eqnarray}
<l|E_i|k> = \epsilon_{ikl}.
\end{eqnarray}
Whereas, for right operators, we have
\begin{eqnarray}
<l|~(1|E_i)~|k> &=& - <k|~(1|E_i)~|l> \q, \\
<0|~(1|E_i)~|k> &=& - <k|~(1|E_i)~|0> = - \delta_{ik} \q, \\
<l|~(1|E_i)~|k> &=& - \epsilon_{ikl} \q .
\end{eqnarray}
We think, it is clear, that these fundamental matrices
are the direct generalizations of 't~Hooft matrices 
\cite{hoft} (the
quaternionic case $i,j,k = 1,2,3$). So, they can play a dual role, for 
a SLA, as we see in this article, and a solitonic construction 
as any 7 or 8 dimensions
instanton is a direct generalization from quaternions
to octonions. Problems arise only
for finding the suitable embedding of $S_7$ 
in a Lie algebra which can be solved directly
by using (\ref{a1}--\ref{a2}) 
and that is all for the time being.

\end{document}